\documentclass[11pt]{article}

\usepackage[final]{acl}

\usepackage{times}
\usepackage{latexsym}

\usepackage[T1]{fontenc}

\usepackage[utf8]{inputenc}

\usepackage{microtype}

\usepackage{inconsolata}

\usepackage{graphicx}

\usepackage{subcaption}
\usepackage{booktabs}
\usepackage{amsmath}
\usepackage{multirow}
\usepackage{makecell}
\usepackage[breakable]{tcolorbox}

\title{From Flat to Hierarchical: Evolving Tree-structured Thoughts\\ for Fine-grained Alpha Mining}

\author{
 \textbf{Junji Ren\textsuperscript},
 \textbf{Junjie Zhao\textsuperscript},
 \textbf{Shengcai Liu\textsuperscript},
 \textbf{Peng Yang\textsuperscript}
}

\begin{document}
\maketitle
\begin{abstract}
Alpha mining, aimed at discovering predictive return signals, is typically formulated as symbolic regression. Traditional symbolic methods suffer from search inefficiency and biased prior knowledge. Recently, Large Language Models (LLMs) have emerged as a promising alternative, automatically generating textual thoughts and executable codes to achieve both efficient and interpretable alpha mining. 
However, existing approaches mostly focus on leveraging LLM's reasoning and reflection capabilities, yet largely neglect the positional bias due to the flat thought representation which restricts efficiency and diversity of the search process. 
This paper introduces Tree-structured thought Evolution (TreEvo), which evolves hierarchically decomposed thoughts to expand the effective search space. In addition, we propose a set of evolutionary operators tailored to structured thoughts. Experiments on four real-market datasets demonstrate that TreEvo not only obtains competitive alphas with traditional methods in up to 200 times fewer evaluations, but also consistently outperforms LLM-driven EAs across all datasets by $14.31\%$ on average.
\end{abstract}

\section{Introduction}

Alpha mining, an important endeavor within financial quantitative trading, is to identify measurable signals that predict asset returns. The core challenge of alpha mining lies in isolating truly predictive factors from noisy, non-stationary data efficiently while ensuring interpretability \cite{giglio2022factor,belle2021principles}. Existing literature primarily tackles this challenge through two distinct computational paradigms. Under the sub-symbolic paradigm, neural networks excel at automatic feature extraction and nonlinear modeling in alpha mining\cite{abe2018deep,chen2024deep,jiang2021applications}, yet their "black-box" nature severely limits interpretability and restricts widespread financial adoption. Conversely, the symbolic paradigm remains predominant because it generates interpretable factors for rigorous strategy validation and risk management\cite{vcernevivciene2024explainable}, though tree-based methods like Genetic Programming (GP)\cite{zhang2020autoalpha} and Reinforcement Learning (RL)\cite{yu2023generating} suffer from an exponentially large search space driven by combinatorial explosion\cite{vie2020qualities}.

\begin{figure}[!t]
  \centering
  \begin{subfigure}[b]{0.45\linewidth}
    \centering
    \includegraphics[width=\linewidth]{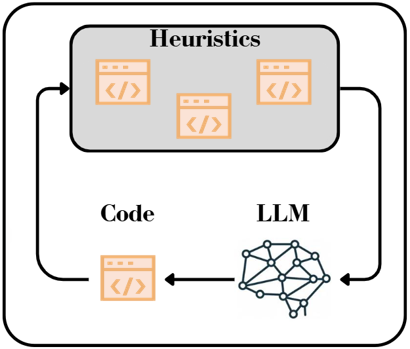}
    \caption{Code (FunSearch)}
    \label{fig:C}
  \end{subfigure}
  \hspace{5pt}
  \begin{subfigure}[b]{0.45\linewidth}
    \centering
    \includegraphics[width=\linewidth]{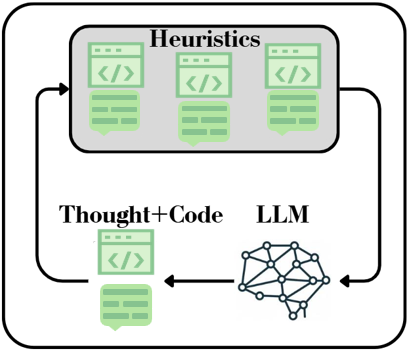}
    \caption{Thought + Code (EoH)}
    \label{fig:TC}
  \end{subfigure}

  \vspace{1em}

  \begin{subfigure}[b]{0.93\linewidth}
    \centering
    \includegraphics[width=\linewidth]{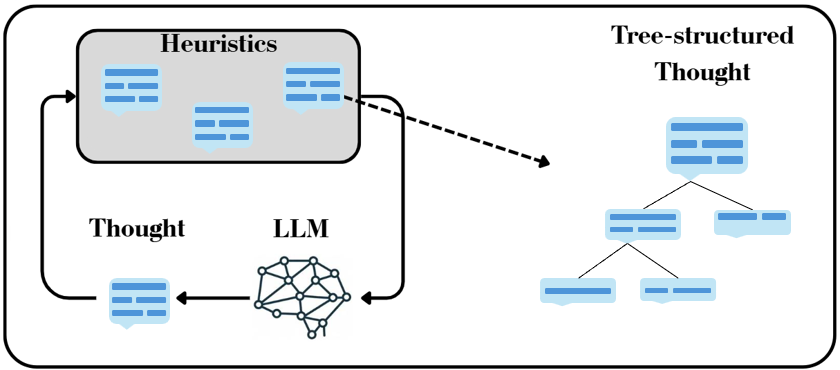}
    \caption{Tree-structured Thought (TreEvo, our work)}
    \label{fig:T}
  \end{subfigure}

  \caption{Different forms of heuristics in LLM-driven EAs.}
  \label{fig:figure_1}
\end{figure}


Recent advances have explored the integration of Large Language Models (LLMs) with Evolutionary Algorithms (EAs)\cite{romera2024mathematical}, which work by iteratively prompting LLMs for evolving the executable codes. Compared to traditional symbolic methods, LLMs provide full expressiveness, under which the search space is no longer limited to manually predefined symbols\cite{merler2024context}. Besides, the massive prior knowledge encoded in LLMs allow for the generation of diverse and context-relevant code structures\cite{jiang2024survey}. Consequently, this leads to faster convergence in possible solutions. More recently, research has expanded to jointly evolve both the underlying thoughts and the generated codes\cite{liu2024evolution}. The shift from code-only to thought-and-code evolution enables more cognitively inspired and effective LLM-driven program synthesis.

Such a framework has been explored to foster a more interpretable and efficient search process of symbolic approaches in alpha mining\cite{li2024can,shi2025navigating}. However, existing methods mostly focus on the enhancement of LLM-based evolutionary operators and trajectories, overlooking the underlying conceptual structure of thoughts. Current LLM-driven methods rely on flat-text thoughts, which constrain the capacity to disentangle logical dependencies. The limitation leads to positional bias alongside redundant refinements, inducing search stagnation and premature convergence. In this context, a transition toward hierarchical thought structures holds potential to alleviate the inherent constraints of linear text.


In this paper, we propose Tree-structured thought Evolution (TreEvo), a novel paradigm where the evolution process is performed on thoughts represented as hierarchical tree structures. Each sub-tree encapsulates a sub-component of reasoning, and higher-level nodes form increasingly abstract heuristics. This structural clarity reshapes the effective topology of the search space, transforming an amorphous, high-dimensional flat-text manifold into a more navigable logical landscape that more closely reflects the recursive structure underlying symbolic discovery. We also propose the tailored evolutionary operators for evolving the hierarchical tree-structure. Empirical comparisons with six representative traditional methods and two LLM-driven EAs have been conducted on stock pools across both Chinese and U.S. markets. TreEvo shows significant advantages over traditional approaches in search efficiency, achieving competitive results with much fewer evaluations. The clear superiority of the proposed tree-structure and evolutionary operators are also extensively analyzed by comparing with two representative LLM-driven EAs and the ablation variants.

In summary, our contributions are as follows:

\begin{itemize}

\item We propose the tree-structured thought, which explicitly organize the reasoning path of an idea in a hierarchical manner to mitigate positional bias in flat text.

\item We propose TreEvo for alpha mining, a novel paradigm where evolution operates at the tree-structured thoughts with the evolutionary operators for fine-grained search.

\item We demonstrate that TreEvo not only achieves competitive performance with traditional methods costing up to 200 times fewer evaluations, but also consistently outperforms LLM-driven EAs on four diverse real-market datasets.

\end{itemize}

\section{Background and Related Works}

\subsection{Alpha Mining}

Alpha mining refers to the automated discovery of predictive signals (alphas) for future asset returns. Early approaches relied on handcrafted designs \cite{fama1993common,grinold2000efficiency}. While recent neural network methods capture complex nonlinear and temporal patterns \cite{heaton2017deep,kim2019financial,qian2024mdgnn}, they often struggle with interpretability and overfitting. Symbolic approaches, using techniques like GP and RL \cite{ren2024alpha,ren2024riskminer,zhao2025quantfactor,crochepierre2022interactive}, generate transparent, modular alpha formulas. Current research focuses on robustness, regime adaptation, and balancing predictive power with interpretability \cite{gu2020empirical,yeo2025comprehensive}, positioning symbolic paradigms as principled alternatives to black-box models \cite{wang2023scientific}.

\subsection{LLM-driven EAs}

The emergence of LLMs has provided novel opportunities to enhance EAs through their advanced capabilities in natural language understanding, symbolic reasoning, and generative modeling \cite{zhang2023survey,romera2024mathematical}. Incorporating implicit knowledge acquired during pretraining, LLMs can support solution generation, guide search trajectories, and enable context-aware evolutionary operators \cite{hao2024large,huang2024exploring,tian2025llm,ye2024reevo}. Recent studies further explore LLM-driven representation learning, program synthesis, and automated operator design to improve search efficiency and reduce reliance on handcrafted heuristics \cite{liu2024large,lange2024large}. Despite challenges related to computational cost, robustness, and interpretability, LLM-augmented EAs represent a rapidly growing direction with strong potential for advancing metaheuristic optimization.

\subsection{Hierarchical Text}

Hierarchical text modeling has evolved from simple sequential processing toward architectures that explicitly capture nested structural dependencies in language. Early studies showed that multi-level representations improve semantic modeling and alleviate semantic dilution in flat architectures \cite{yang2016hierarchical}. Subsequent work demonstrated that hierarchical structures enhance long-document representation and content aggregation beyond the capacity of standard models \cite{cohan-etal-2018-discourse,zhang-etal-2019-hibert}. More recent research highlights their role in recursive reasoning, knowledge navigation, and interpretability in complex information settings \cite{he-etal-2025-hmt,huang-etal-2025-retrieval}. However, existing studies primarily focus on hierarchical text classification, whereas comparatively limited attention has been given to treating hierarchical text itself as a structured search space for LLMs.

\section{Method}

\subsection{Main Idea}

TreEvo encodes each candidate alpha represented as a tree \( \mathcal{T} = (V, E) \), where \( V \) is the set of reasoning units (e.g., logical steps), and \( E \subseteq V \times V \) defines the hierarchical dependencies between them. Each node \( v_i \in V \), corresponds to a distinct semantic or computational sub-component, and the root node \( v_0 \) encodes the expression of the overall alpha factor. The hierarchy reflects the structural organization of human cognition, which is widely understood to proceed from abstract, global concepts toward more concrete, fine-grained elaborations. 

The hierarchical structure induces a structured parameterization of the search space $\mathcal{S} = \{\mathcal{T}_1, \mathcal{T}_2, \ldots\}$ by mapping candidate alphas onto a compositional tree topology $\mathcal{T}$, where structural variation is formally expressed as a set of transformation operators $\Phi \colon \mathcal{T} \to \mathcal{T}'$ over sub-trees rather than modifications to a monolithic sequence. This topology partitions the search space into weakly coupled regions $\mathcal{S} = \bigcup_k \mathcal{S}_k$ associated with different functional roles, which constrains how changes propagate across the representation. As a result, semantic dependencies are mediated through explicit structural interfaces instead of implicit global entanglement. The search process is therefore organized around operations on bounded structural units with controllable scopes, forming a landscape $\mathcal{L} = (\mathcal{S}, \Phi)$ in which alternative configurations arise through systematic recombination and rearrangement of these units.

\subsection{Tree-structured Thought Representation}

\begin{figure}[t]
    \centering
    \begin{subfigure}{\linewidth}
        \centering
        \includegraphics[width=\linewidth]{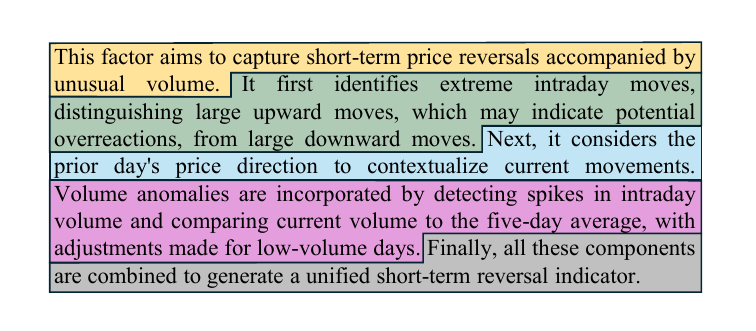}
        \caption{Flat text}
        \label{fig:flat_text}
    \end{subfigure}

    \vspace{0.3cm}

    \begin{subfigure}{\linewidth}
        \centering
        \includegraphics[width=\linewidth]{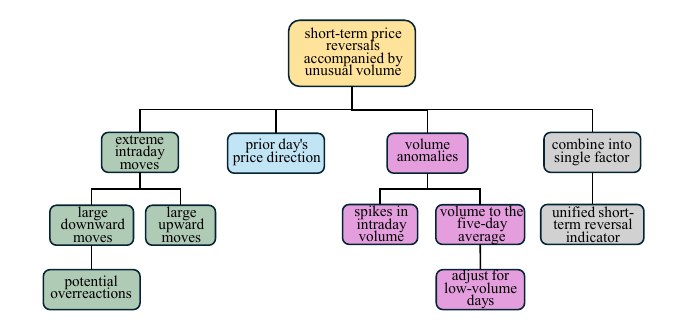}
        \caption{Tree-structured text}
        \label{fig:tree_text}
    \end{subfigure}

    \caption{An example of a set of paired flat text and tree-structured text.}
    \label{fig:tree_flat}
\end{figure}

Figure \ref{fig:tree_flat} shows an example of a flat text and its corresponding hierarchical tree representation used to organize complex concepts or computations. The sentences with the same color in \ref{fig:flat_text} and \ref{fig:tree_text} correspond to each other. At the highest level of the tree-structured text, a root node represents the overall entity or idea, which is decomposed into several child nodes capturing its main reasoning steps. Each child node can further branch into its own children, reflecting more finer-grained operations that collectively define the meaning of their parent node. This layer-wise decomposition continues recursively until the leaf nodes that represent the most basic, indivisible units in the structure. The edges between nodes indicate the compositional relationships, where parent nodes depend on or aggregate information from their children. The hierarchical levels correspond to different degrees of abstraction, with higher levels summarizing or coordinating the details specified at lower levels.

In TreEvo, the entire tree-structured thought is generated by LLM as a holistic structured entity, with hierarchical information explicitly encoded using structured symbolic delimiters. Each node and its children are separated by these symbols, establishing clear boundaries that distinguish reasoning units at different levels of abstraction. This allows the model to produce a complete hierarchical representation in one step, while preserving the modularity and compositionality of its constituent sub-trees. By maintaining these explicit structural cues, the generated thought provides a coherent framework that can later be systematically explored and modified, supporting principled evolution of alternative tree configurations.

On this basis, TreEvo is significantly different from the well-known Chain-of-Thought(CoT) \cite{wei2022chain} and Tree-of-Thought(ToT) \cite{yao2023tree}. CoT treats a thought as a sequence of reasoning steps. ToT also organizes a thought as a sequence of reasoning steps, and integrates multiple related thoughts as a tree. That is, sibling reasoning steps under the same sub-tree typically share the same reasoning prefix. In contrast, TreEvo regards a whole tree as a thought. Consequently, the generation of the thoughts are also different. In CoT, the whole thought is generated by constructively appending reasoning steps. In ToT, the generation of a thought is to find the optimal complete branch from the root to one leaf. In TreEvo, we need to search the space of tree-structured representations, where EA can be adopted.

\subsection{Evolutionary Operators}

We propose five evolutionary operators to generate new individuals in the search process. Collectively, they are designed to balance recombination, variation, and simplification within the tree-structured search space. The crossover operator primarily emphasizes recombination of coherent reasoning modules, facilitating the emergence of new structural compositions. The mutation operators focus on introducing controlled variations at different levels, supporting both global exploration and local refinement of candidate structures. The pruning operator prioritizes structural simplification, encouraging more compact and parsimonious representations while maintaining functional validity.

    \noindent\textbf{Crossover}:
    Given two individuals \( \mathcal{T}_1 \) and \( \mathcal{T}_2 \), we define crossover as integrating sub-trees:
    \[
    \mathcal{T}' = \mathcal{T}_1 \oplus_{v_i, v_j} \mathcal{T}_2
    \]
    where \( v_i \in \mathcal{T}_1 \), \( v_j \in \mathcal{T}_2 \). \( \oplus \) denotes the integration of \( v_i\) and \( v_j \). This operator preserves semantic validity and allows meaningful recombination of reasoning units.

    \noindent\textbf{Mutation-R}:
    It generates an individual by generating a completely different tree \( \widetilde{\mathcal{T}} \) from current population \( \{\mathcal{T}_1,...,\mathcal{T}_n\} \):
    \[
    \mathcal{T}' = \widetilde{\mathcal{T}}
    \]
    This ensures the diversity of individuals in the population.

    \noindent\textbf{Mutation-I}:
    It operates \( \mathcal{T} \) by selecting an internal node \( {v_k} \) and replacing its sub-tree with a newly generated, syntactically valid and semantically plausible tree \( \widetilde{\mathcal{T}} \):
    \[
    \mathcal{T}' = \mathcal{T} \ominus_{v_k} \widetilde{\mathcal{T}}
    \]
    where \( \ominus \) denotes the sub-tree or node replacement rooted at node \( {v_k} \). This allows localized alteration of logical steps while preserving the overall structure and intent of the thought.
    
    \noindent\textbf{Mutation-F}:
    It operates \( \mathcal{T} \) by selecting a leaf node \( {v_k} \) and replacing it with a newly generated, syntactically valid and semantically plausible node \( v' \):
    \[
    \mathcal{T}' = \mathcal{T} \ominus_{v_k} v'
    \]
    This allows refinements on parameter only while preserving the overall structure and intent of the thought.
    
    \noindent\textbf{Pruning}:
    It operates by identifying and removing structurally redundant or semantically neutral sub-trees from \( \mathcal{T} \). Given the sub-tree of node \( v_k \), if it satisfies the redundancy criteria, it is replaced with a simpler expression \( \widehat{\mathcal{T}} \):
    \[
    \mathcal{T}' = \mathcal{T} \ominus_{v_k} \widehat{\mathcal{T}}
    \]
    This facilitates structural simplification and encourages parsimony in the thought representation.

All evolutionary operators are implemented at the prompt level with explicit constraints that preserve the logical and semantic coherence of individuals. During each operation, LLM is guided to maintain structural consistency and respect the hierarchical dependencies within the tree representation. As a result, newly generated individuals remain semantically valid and logically self-consistent, ensuring that evolutionary modifications operate on well-formed reasoning structures rather than arbitrary textual variations.

\subsection{Framework}

The evolution process maintains a population of $N$ candidate alphas, which are represented as tree-structured thoughts. The population are iteratively evolved with LLM. At each iteration, given $N$ parent thoughts, $N$ new offspring thoughts are generated by LLM with one of three operator types: crossover operator, mutation operators, or pruning operator. To evaluate those $N$ new thoughts, LLM is first applied to generate one piece of executable code based on each new thought, then each code is run with the given market dataset to calculate the predictive performance. After that, a new population of $N$ candidate thoughts are selected among the $N$ parents and $N$ offspring, based on their individual evaluated performance. More specific steps are described in Appendix \ref{app:prompt}.

\section{Experiments}

The experimental studies focus on three key research questions (RQs):

\begin{itemize}
    \item \noindent\textbf{RQ1:}
    How tree-structure contributes to LLM-driven EAs for fine-grained search process? 

    \item \noindent\textbf{RQ2:}
    How does our proposed method compared to existing methods on alpha mining?

    \item \noindent\textbf{RQ3:}
    Can the idea of tree-structure also be applied to existing LLM-driven EAs? How does the evolutionary operators contribute?
\end{itemize}

\subsection{Experimental Settings}

\subsubsection{Dataset}

Four widely used stock pools that serve as benchmarks are considered, encompassing both Chinese A-shares market (CSI300, CSI500) and U.S. stock market (SPX, NDX). Considering about the reproducibility, we chose six widely used raw data features, opening, closing, highest, lowest prices (OHLC), trading volume (volume), and volume-weighted average price (vwap), for the whole experiments. All price and volume data are forward-adjusted for corporate actions to ensure consistency over time. The dataset is split chronologically into training set (2016/01/01-2020/01/01), validation set (2020/01/01-2021/01/01) and test set (2021/01/01-2024/01/01). 

\subsubsection{Baselines} The framework is compared with eight baseline methods. Six of them are traditional approaches: \textbf{XGBoost} \cite{chen2016xgboost}, \textbf{LightGBM} \cite{ke2017lightgbm}, \textbf{GP} \cite{zhang2020autoalpha}, \textbf{AlphaGen} \cite{yu2023generating}, \textbf{AlphaForge} \cite{shi2025alphaforge} and \textbf{QFR} \cite{zhao2025quantfactor}. The remaining two are LLM-driven EAs for code generation: \textbf{EoH} \cite{liu2024evolution} and \textbf{ReEvo} \cite{ye2024reevo}. 22 arithmetic operators are used for combing the above six data features in traditional approaches (see Appendix \ref{app:operator}), while no restriction for the operators of LLM-driven EAs given the embedded prior knowledge.

For traditional approaches, we adopt the evaluation budgets specified in their original implementations. The hyperparameters are consistent with their original papers. Specifically, the hyperparameter settings of XGBoost and LightGBM follow those provided in Qlib\cite{yang2020qlib}, which consume 1000 evaluations. GP adopts the default hyperparameter configuration in gplearn, which consumes 40000 evaluations. AlphaGen, AlphaForge and QFR follow the configurations reported in their original papers, which consume at least 50000 evaluations. For LLM-driven EAs, we set the evaluation budget for obtaining the best alpha to a small number of 200 evaluations. Models utilize the Qwen3-Max model \cite{yang2025qwen3} to generate both thoughts and codes of candidate solutions, and the population size $N$ is fixed to 10. For TreEvo, $p_R$ and $p_I$ is set to 0.4, $p_F$ is set to 0.2. The performance of an algorithm on each testing dataset is calculated by averaging over 5 independent runs of the above process, to ensure fairness and consistency.

\subsubsection{Evaluation Metrics}

To measure the predictive performance of each algorithm, we employ two metrics commonly used across the quantitative finance industries. IC measures the cross-sectional Pearson correlation between factor values $F \in \mathbf{R}^{n \times T}$ and the corresponding future 5-day returns $R \in \mathbf{R}^{n \times T}$ over $T$ trading days and $n$ stocks as

\begin{align}
    &IC(F,R) = \frac{1}{T}\sum_{t=1}^T corr(f_t, r_t) \\
    &= \frac{1}{T}\sum_{t=1}^T \frac{\sum_{i=1}^{n}(f_{i,t} - \overline{f}_{t})(r_{i,t} - \overline{r}_{t})}{\sqrt{\sum_{i=1}^{n}(f_{i,t} - \overline{f}_{t})^2\sum_{i=1}^{n}(r_{i,t} - \overline{r}_{t})^2}}.
\end{align}

\noindent It evaluates the predictive power of a mined alpha with values ranging from $-1$ to $1$, and is also adopted as the training loss or search objective for the compared methods. RankIC is defined as the Spearman rank correlation between $\mathbf{F}$ and $\mathbf{R}$ as 

\begin{align}
    RankIC(F,R) = IC(rank(F), rank(R)),
\end{align}

\noindent where $rank()$ assigns ranks to the elements of $F$ or $R$ each day across all stocks, capturing monotonic relationships and exhibiting greater robustness to outliers and scale variations.

\subsection{Results and Analysis}

\subsubsection{Comparisons of Text Representations}

To illustrate the advantage of tree-structured text in enabling fine-grained search within LLM-driven evolutionary algorithms, we conducted experiments to examine how the LLM applies changes to tree-structured versus flat text. As a first experiment, we generated 10 sets of paired samples with both formats, each consisting of 10 sub-ideas, and use a simple prompt to select a single sub-idea as the variation point for generating a new individual. To eliminate order bias, half of the sets were generated starting with tree-structured text and the other half with flat text (details in Appendix \ref{app:variation}). This setup allows a direct comparison of the modification flexibility and locality afforded by each representation.

Figure \ref{fig:t2f1x5} presents the distribution of variation points of both forms of text. The distribution of variation points indicates a notable distinction between tree-structured and flat text representations. In tree-structured text, ideas are selected relatively uniformly, suggesting that the model can exploit a wide range of positions for modification, thereby facilitating fine-grained and diverse search. In contrast, flat text exhibits an uneven distribution, with certain positions being preferentially selected while others are seldom modified, implying a bias toward localized changes and restricted exploration. Consistent with the observations, the tree-structured representation achieves a mean MSE of 0.00869 relative to the uniform distribution, whereas the flat-text representation records a higher value of 0.01829, corresponding to an approximate $52.49\%$ reduction.

\begin{figure}[t]
     \centering
     \begin{subfigure}[b]{0.23\textwidth}
         \centering
         \includegraphics[width=\textwidth]{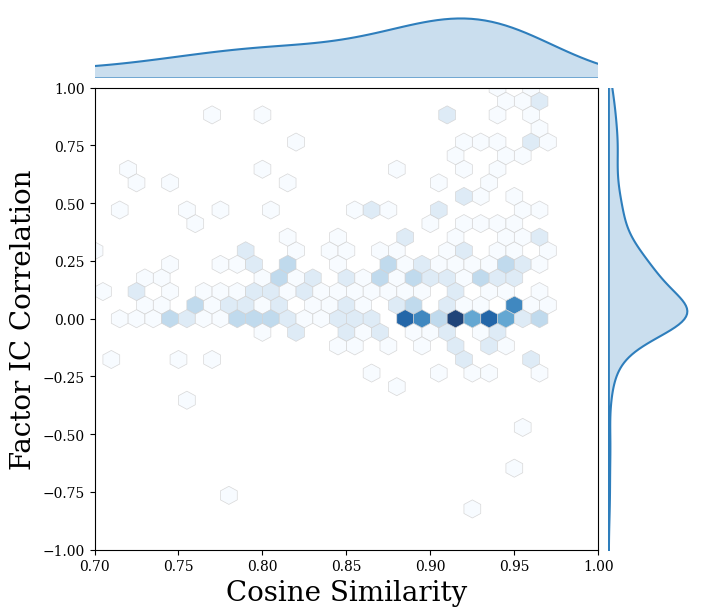}
         \caption{Tree-structured Text}
         \label{fig:cosine_ic_tree}
     \end{subfigure}
     \begin{subfigure}[b]{0.23\textwidth}
         \centering
         \includegraphics[width=\textwidth]{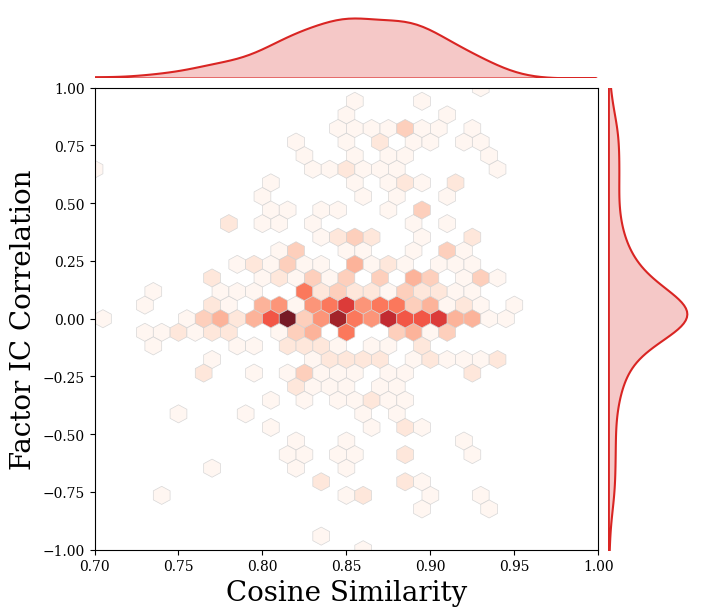}
         \caption{Flat Text}
         \label{fig:cosine_ic_flat}
     \end{subfigure}
     
     \caption{Comparison between distribution of textual cosine similarity and factor IC correlation in tree-structured and flat text formats.}
     \label{fig:distribution}
\end{figure}

We further examine how the tree-structured representation contributes to the optimization characteristics of the search space. Specifically, we focus on the relationship between textual cosine similarity and factor IC correlation of the paired texts. For each pair, we initially utilized a prompt to randomly generate a base factor text under a given representation form. Then, we employed the identical prompt to instruct the LLM to apply a modification to this base text, thereby producing its corresponding variant (details in Appendix \ref{app:simi_corr}).

As shown in Figure \ref{fig:distribution}, the joint distribution of textual cosine similarity and factor IC correlation illustrates distinct patterns between the two representation forms. In the flat text, the similarities are highly clustered within a narrow textual similarity range of 0.8 to 0.9, while the corresponding factor IC correlations scatter widely across the entire range from -1 to 1. In contrast, the tree-structured text exhibits a significantly broader and more uniform distribution of textual similarity, spanning from 0.7 to 0.95 with a flatter marginal density curve. Moreover, the vast majority of paired texts maintain positive factor IC correlations, showing a more stable distribution with fewer extreme negative anomalies. This indicates that the tree-structured representation effectively smooths the optimization landscape, ensuring that structural variations yield more stable and directional search.

\subsubsection{Comparisons with Traditional Methods}
Table \ref{tab:performance} shows the performance comparisons of traditional approaches and TreEvo on CSI300 and CSI500. Overall, TreEvo shows comparable performance to the baselines in terms of IC, illustrating its competitiveness in capturing the magnitude of future returns. On CSI300, TreEvo’s delivers an IC improvement of over $1.99\%$ over the strongest baseline. On CSI500, TreEvo remains in the top tier, surpassing all baseline models in IC except for QFR. These results highlight the benefit of incorporating LLM guidance into the alpha mining.

\begin{table}[t]
\scriptsize
\centering
\caption{Comparisons of IC and RankIC with Traditional Approaches}
\label{tab:performance}
\begin{tabular*}{\linewidth}{@{\extracolsep{\fill}}lcccc}
\toprule
 Method & \multicolumn{2}{c}{\textbf{CSI300}} & \multicolumn{2}{c}{\textbf{CSI500}} \\
\cmidrule(lr){2-3} \cmidrule(lr){4-5}
 & \textbf{IC} & \textbf{RankIC} & \textbf{IC} & \textbf{RankIC} \\
\midrule
XGBoost & 0.0197 & 0.0221 & 0.0169 & 0.0209 \\
 & ($\pm$0.0025) & ($\pm$0.0013) & ($\pm$0.0016) & ($\pm$0.0021) \\
\addlinespace
LightGBM & 0.0165 & 0.0244 & 0.0172 & 0.0241 \\
 & ($\pm$0.0012) & ($\pm$0.0025) & ($\pm$0.0022) & ($\pm$0.0017) \\
\addlinespace
GP & 0.0224 & 0.0278 & 0.0254 & 0.0297 \\
 & ($\pm$0.0017) & ($\pm$0.0038) & ($\pm$0.0028) & ($\pm$0.0014) \\
\addlinespace
AlphaGen & 0.0259 & 0.0301 & 0.0337 & \textbf{0.0405} \\
 & ($\pm$0.0010) & ($\pm$0.0046) & ($\pm$0.0023) & (\textbf{$\pm$0.0047}) \\
\addlinespace
AlphaForge & 0.0287 & \textbf{0.0351} & 0.0319 & 0.0346 \\
 & ($\pm$0.0014) & (\textbf{$\pm$0.0029}) & ($\pm$0.0022) & ($\pm$0.0026) \\
\addlinespace
QFR & \textbf{0.0302} & 0.0346 & \textbf{0.0374} & 0.0403 \\
 & (\textbf{$\pm$0.0032}) & ($\pm$0.0044) & (\textbf{$\pm$0.0039}) & ($\pm$0.0031) \\
\midrule
TreEvo & \textbf{0.0308} & \textbf{0.0349} & \textbf{0.0362} & \textbf{0.0393} \\
 & (\textbf{$\pm$0.0048}) & (\textbf{$\pm$0.0057}) & (\textbf{$\pm$0.0051}) & \textbf{($\pm$0.0049)} \\
\bottomrule
\end{tabular*}
\end{table}

\begin{figure}[t]
    \centering
    \includegraphics[width=0.45\textwidth]{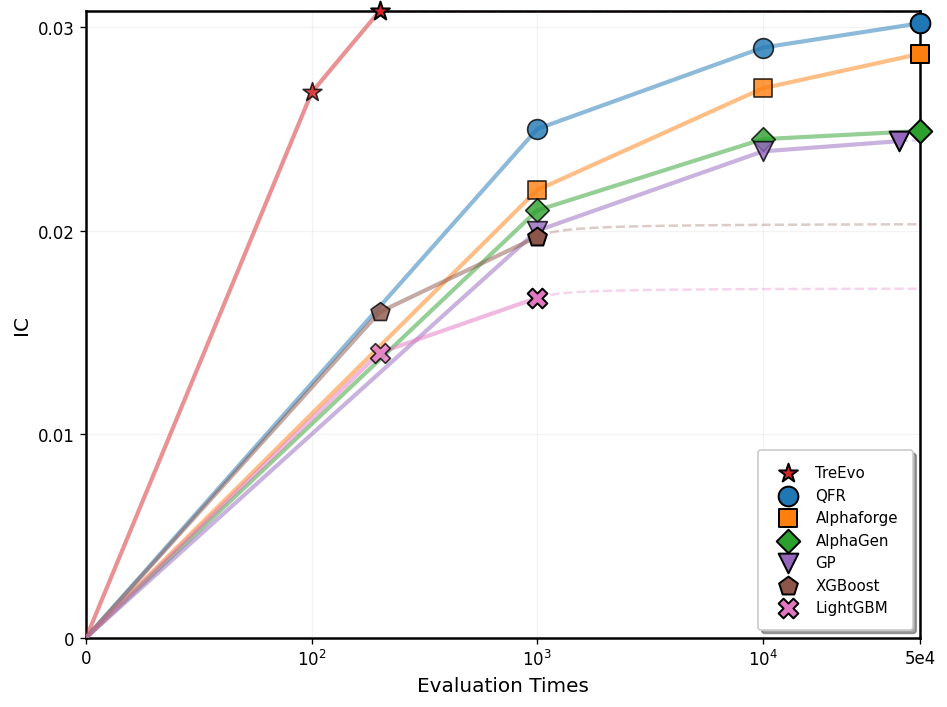}
    \caption{Comparison of search efficiency with Traditional Approaches on CSI300.}
    \label{fig:eval_ic}
\end{figure}

Figure \ref{fig:eval_ic} indicates the high efficiency of TreEvo in alpha mining compared to traditional approaches. TreEvo achieves higher IC within 5 times fewer evaluations than XGBoost and LightGBM and up to 200 times fewer than the remaining traditional approaches. This efficiency underscores the effectiveness of LLM-driven search methods in narrowing the search space with the embedded prior knowledge and facilitating faster convergence toward high-quality solutions. The reduced resource demands, combined with improved predictive accuracy, position TreEvo as a highly practical and worth exploring framework for alpha mining. Wall-clock time and token costs are shown in Appendix \ref{app:clock}.

In Figure \ref{fig:cumulative_reward}, we further depict the cumulative return of each method on the CSI300 index with the commonly used Top-50/Drop-5 trading strategy. The trading strategy selects the top 50 stocks based on their alpha scores with a daily rebalancing frequency. At each rebalance date, the bottom five stocks in the current portfolio are removed and replaced with five new candidates from the updated top-ranked stocks. Figure \ref{fig:cumulative_reward} demonstrates the superiority of TreEvo over the traditional approaches in terms of cumulative return under this strategy. Except for the very short initial period, TreEvo maintains the leading position throughout the backtest. On the whole, TreEvo achieves $21.17\%$ in the cumulative return and more than $58.42\%$ excess return by compared to the CSI300 market benchmark (i.e., the gray dash curve in the figure). Moreover, as the CSI300 index continuously goes down, the increasing curve of TreEvo demonstrates that real alphas that are less correlated to beta have been mined.

To conclude, the performance of TreEvo not only shows its superiority over representative approaches in the quantitative finance industries, but also supports that LLM-driven EAs can be revolutionary tools for alpha mining. LLM-driven EAs can automatically generate executable codes of powerful alphas in minutes rather than traditionally daily, and without much human efforts in the manufacturing loop.

\begin{figure}[t]
    \centering
    \includegraphics[width=0.48\textwidth]{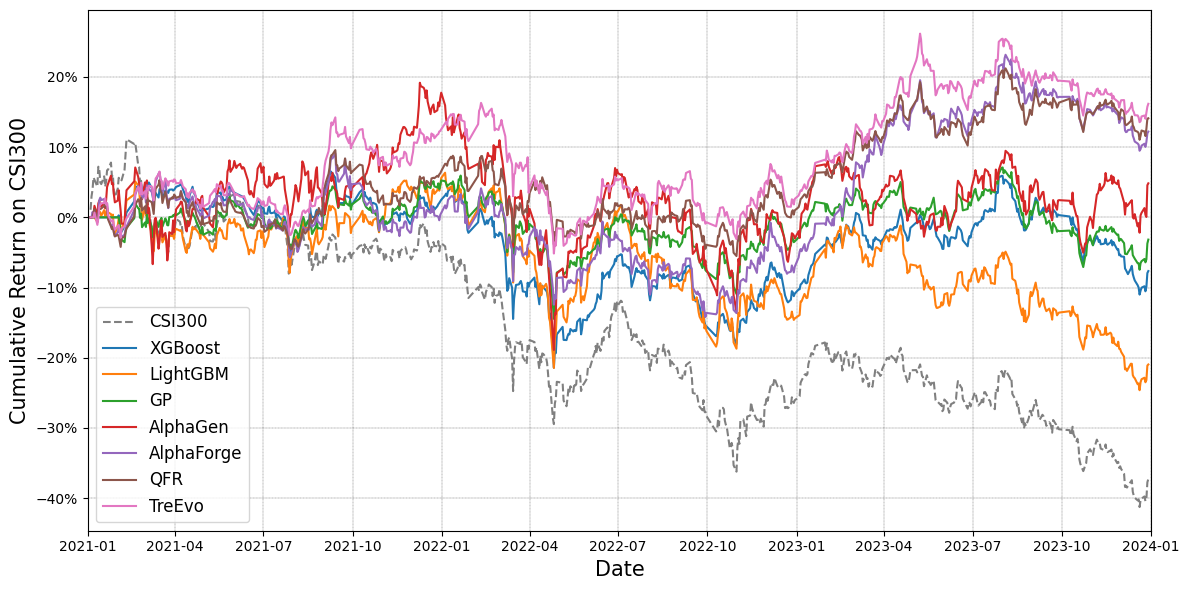}
    \caption{Cumulative rewards of traditional approaches on CSI300.}
    \label{fig:cumulative_reward}
\end{figure}

\subsubsection{Comparisons with LLM-driven EAs} Table \ref{tab:A_performance} reports the comparative results of TreEvo against two existing LLM-driven EAs across all datasets. The results demonstrate that TreEvo generally outperforms the compared approaches, indicating that the alpha signals discovered by our method possess stronger linear associations with future returns across different markets. Overall, TreEvo outperforms the strongest baseline by an average of $14.31\%$. More specifically, on CSI300, SPX and NDX, the improvement made by TreEvo on IC is over $10\%$ , which is significant. Notably, it is observed that EoH consistently beats ReEvo on all cases. This shows that evolving thoughts and codes separately is better than solely evolving the codes. Moreover, given TreEvo outperforms EoH, this table further suggests that evolving tree-structured thoughts is more effective than plain-text thoughts, which verifies the major hypothesis of this work.

\begin{table}[t] 
\scriptsize
\centering
\caption{Performance Comparisons of LLM-driven EAs}
\label{tab:A_performance}
\begin{tabular*}{\linewidth}{@{\extracolsep{\fill}}llccc}
\toprule
\textbf{Dataset} & \textbf{Metric} & \textbf{EoH} & \textbf{ReEvo} & \textbf{TreEvo} \\
\midrule
\multirow{2}{*}[-1.2ex]{CSI300} & IC & \makecell{0.0238 \\ ($\pm$0.0028)} & \makecell{0.0211 \\ ($\pm$0.0017)} & \makecell{\textbf{0.0308} \\ \textbf{($\pm$0.0033)}} \\
 & RankIC & \makecell{0.0262 \\ ($\pm$0.0037)} & \makecell{0.0269 \\ ($\pm$0.0028)} & \makecell{\textbf{0.0349} \\ \textbf{($\pm$0.0057)}} \\
\midrule
\multirow{2}{*}[-1.2ex]{CSI500} & IC & \makecell{0.0348 \\ ($\pm$0.0036)} & \makecell{0.0247 \\ ($\pm$0.0031)} & \makecell{\textbf{0.0362} \\ \textbf{($\pm$0.0041)}} \\
 & RankIC & \makecell{0.0363 \\ ($\pm$0.0046)} & \makecell{0.0283 \\ ($\pm$0.0042)} & \makecell{\textbf{0.0393} \\ \textbf{($\pm$0.0049)}} \\
\midrule
\multirow{2}{*}[-1.2ex]{SPX} & IC & \makecell{0.0284 \\ ($\pm$0.0026)} & \makecell{0.0224 \\ ($\pm$0.0033)} & \makecell{\textbf{0.0317} \\ \textbf{($\pm$0.0035)}} \\
 & RankIC & \makecell{0.0326 \\ ($\pm$0.0030)} & \makecell{0.0268 \\ ($\pm$0.0036)} & \makecell{\textbf{0.0355} \\ \textbf{($\pm$0.0037)}} \\
\midrule
\multirow{2}{*}[-1.2ex]{NDX} & IC & \makecell{0.0254 \\ ($\pm$0.0029)} & \makecell{0.0252 \\ ($\pm$0.0024)} & \makecell{\textbf{0.0285} \\ \textbf{($\pm$0.0031)}} \\
 & RankIC & \makecell{0.0297 \\ ($\pm$0.0034)} & \makecell{0.0302 \\ ($\pm$0.0027)} & \makecell{\textbf{0.0316} \\ \textbf{($\pm$0.0043)}} \\
\bottomrule
\end{tabular*}
\end{table}

We analyze the correlation among the best factors discovered by each method (see Figure \ref{fig:corr}). The heatmaps reveal that ReEvo suffers from significant redundancy, characterized by consistently high positive correlations, while EoH exhibits extreme absolute correlations in both directions, suggesting that its factors are largely interdependent or simple linear transformations of one another. In sharp contrast, TreEvo yields a much sparser correlation profile with significantly lower off-diagonal coefficients. This comparison demonstrates that TreEvo effectively explores a broader functional space and captures more independent alpha signals, thereby mitigating the risk of information overlap and providing a more robust foundation for factor ensemble and portfolio construction. Additionally, we compared TreEvo with existing LLM-based alpha mining frameworks and evaluated its performance across different LLMs (see Appendix \ref{app:llm_frame} and \ref{app:llm_perform}). We also conducted walk-forward validation to evaluate the the out-of-sample performance of the proposed method (see Appendix \ref{app:forward}).

\subsubsection{Ablation Studies}

To clarify whether the major performance gain stems from the tree-structured thoughts or the proposed evolutionary operators, a variant of ReEvo is designed by solely evolving its codes with the proposed tree-structured thoughts, denoted as TReEvo. By comparing TReEvo with ReEvo, it suffices to show how tree-structure contributes to the performance, as they share the same evolutionary operators. By comparing TReEvo with the proposed TreEvo, the novel evolutionary operators are ablated.   

\begin{figure}[t]
    \centering
    \includegraphics[width=0.45\textwidth]{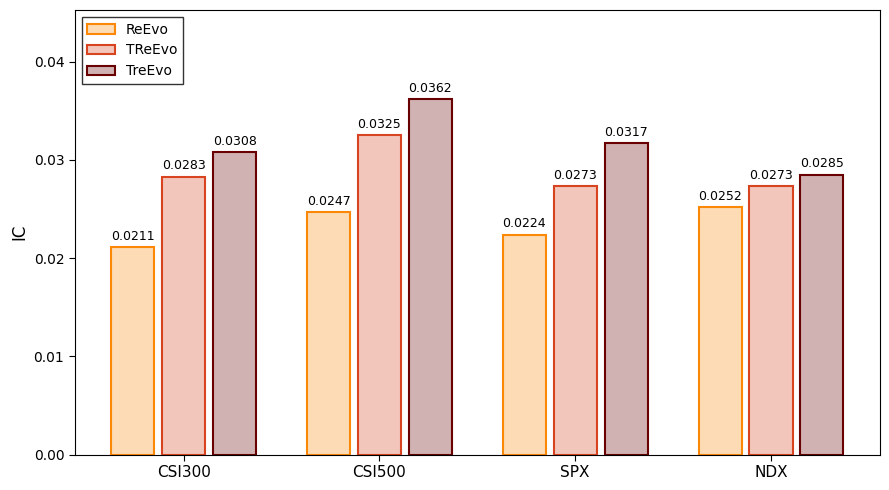} 
    \caption{Comparison results of ablation studies of ReEvo, TReEvo and TreEvo on 4 databases}
    \label{fig:ablation}
\end{figure}

Figure~\ref{fig:ablation} depicts the performance of ReEvo, TReEvo, and TreEvo on four datasets. TReEvo consistently outperforms ReEvo in terms of IC, verifying the advantages of the tree-structured thoughts. The superiority of TreEvo over TReEvo reveals that the proposed evolutionary operators are more compatible with tree-structured thoughts. More specifically, the ICs of TReEvo lead $23.98\%$ over that of ReEvo on average. Comparatively, the ICs of the proposed TreEvo gain advantage over TReEvo by $10.19\%$. These present that the tree-structured thought is more dominant than the novel evolutionary operators to improve the performance of LLM-driven EAs on alpha mining. Meanwhile, the evolutionary operators should also be tailored to be compatible with the hierarchical thoughts, otherwise the improvements would be limited.

\section{Conclusion}

This paper proposes a novel LLM-driven alpha mining paradigm based on tree-structured thoughts, which enables more expressive and hierarchical representations of heuristics. By leveraging the compositional nature of trees, our approach reduces semantic coupling between sub-components, allowing for finer-grained reasoning and more flexible manipulation of intermediate thought steps, making it particularly suitable for complex search spaces such as formulaic alpha. Experiments show that, under a relatively small evaluation budget, our method achieves performance competitive with traditional approaches. The proposed method also demonstrates stronger performance compared with existing LLM-driven EAs under then same evaluation budget.

\section*{Limitations}

An important aspect of TreEvo concerns the explicit hierarchical text representation employed in this work. While the framework adopts a structured encoding with explicit delimiters to express tree topology, we do not systematically examine which forms of hierarchical representation are most effective. Different encoding schemes may introduce distinct inductive biases into the evolutionary process, particularly in how operators enable fine-grained structural control. The choice of representation therefore has meaningful implications for the precision and interpretability of evolutionary search. Investigating alternative hierarchical encodings and their influence on fine-grained search behavior represents a promising direction for further study.

In practical deployments, certain obstacles remain. LLM queries limit the number of individuals that can be generated within a fixed time budget, leading LLM-driven EAs to typically explore fewer candidates than traditional methods. According to experimental observations, mutation operations under the hierarchical representation tend to produce incremental structural updates. Pruning partially alleviates this effect, though introducing more diverse transformations during mutation should further improve exploration. In addition, for constrained tasks such as alpha mining, invalid outputs may still occur when domain constraints are explicitly specified, necessitating additional validation and correction.


\bibliography{custom}

\appendix

\section{Datasets}
\label{app:dataset}

\noindent\textbf{CSI300} tracks the 300 largest and most liquid A-share stocks, which represents the large-capitalization (large-cap) segment of China's market.

\noindent\textbf{CSI500} consists of 500 mid-cap and small-cap A-share stocks, which reflects the behaviors of a broader market besides the large-cap stocks, capturing the growth potential of more dynamic and innovative stocks.

\noindent\textbf{SPX} (S$\&$P500) includes 500 leading publicly traded companies in U.S., covering all major industries and reflecting the overall American economic.

\noindent\textbf{NDX} (NASDAQ100) consists of 100 largest non-financial companies listed on the NASDAQ stock exchange, representing a broad spectrum of technology, consumer, and healthcare sectors in the U.S. economy.

\section{Baselines}
\label{app:baseilne}

\noindent\textbf{Traditional Approaches:} \textbf{XGBoost} \cite{chen2016xgboost} is an implementation of gradient boosted decision trees with regularization and second-order optimization for accurate and scalable prediction on tabular data, provided by Qlib \cite{yang2020qlibaiorient}. \textbf{LightGBM} \cite{ke2017lightgbm} is a gradient boosting framework using histogram-based binning and leaf-wise tree growth for fast and efficient training on large datasets, also provided by Qlib. \textbf{GP} \cite{zhang2020autoalpha} is an EA, iteratively searching interpretable symbolic formulas through evolutionary operators like crossover and mutation, integrated by gplearn\footnote{https://github.com/trevorstephens/gplearn}. \textbf{AlphaGen} \cite{yu2023generating} is an RL framework where PPO‐based RL and tokenized Reverse Polish Notation formula generation optimize portfolio‑level performance to produce synergistic formulaic alpha sets. \textbf{AlphaForge} \cite{shi2025alphaforge} is an RL framework combines a generative‑predictive neural network for formulaic alpha factor mining with a dynamic factor combination model. \textbf{QFR} \cite{zhao2025quantfactor} is an RL approach uses a REINFORCE policy without critic, introducing greedy‑baseline variance reduction and information‑ratio based reward shaping to mine stable, interpretable alpha factors.

\noindent\textbf{LLM-driven EAs:} \textbf{EoH} \cite{liu2024evolution} is a framework that combines LLMs and EA to evolve both thoughts and executable code, enabling efficient automatic heuristic design. \textbf{ReEvo} \cite{ye2024reevo} integrates LLMs with EA to solely generate executable codes, strengthened with reflective reasoning.

\section{Arithmetic Operators}
\label{app:operator}

The predefined operators used in traditional symbolic approaches are listed in Table \ref{tab:operators}.

\begin{table}[htbp]
\footnotesize
\centering
\caption{Operators Used in Symbolic Approaches}
\label{tab:operators}
\begin{tabular}{lc}
\hline
\textbf{Category} & \textbf{Operator} \\ \toprule
\multirow{2}{*}{Cross-Section} & abs($x$), log($x$) \\
            & $+, -, \times, /, >, <$ \\ \midrule
\multirow{6}{*}{Time-Series} & ts$\_$mean($x, l$), ts$\_$med($x, l$), ts$\_$sum($x, l$) \\
            & ts$\_$std($x, l$), ts$\_$var($x, l$) \\
            & ts$\_$max($x, l$), ts$\_$min($x, l$) \\
            & ref($x, l$), mad($x, l$), delta($x, l$) \\
            & wma($x, l$), ema($x, l$) \\
            & ts$\_$cov($x, y, l$), ts$\_$corr($x, y, l$) \\ \bottomrule
\end{tabular}
\end{table}

\section{Variation Point Modification}
\label{app:variation}

We generated paired samples consisting of tree-structured and flat texts containing 10 ideas. These ideas encompass a hierarchy ranging from high-level abstractions to concrete details. For each paired sample, a simple prompt was used to instruct the LLM to select a single idea as a variation point, thereby generating a new individual. The experiment was conducted across 10 independent sets of samples: half of them were generated as tree-structured text first then flat text, the other half were flat text first then tree-structured text.

\section{Alignment Between Textual Similarity and Factor Correlation}
\label{app:simi_corr} 

For each representation, we generated paired samples consisting of two texts containing an initial text and a corresponding variant generated from it. The initial text is produced by an LLM under a specified representation, and the variant is obtained by prompting the LLM to modify the original text under the same prompt, resulting in a structurally related counterpart. Textual similarity is defined as the cosine similarity between embeddings of the two texts. Factor correlation is measured as the IC between the corresponding factors, defined as the correlation between their factor value time series.

\section{Additional Experimental Results}
\label{app:result}

\subsection{Variation Point Distribution}
\label{app:dis}

\begin{figure*}[htbp]
    \centering
    \begin{subfigure}[b]{0.245\textwidth}
        \centering
        \includegraphics[width=\textwidth]{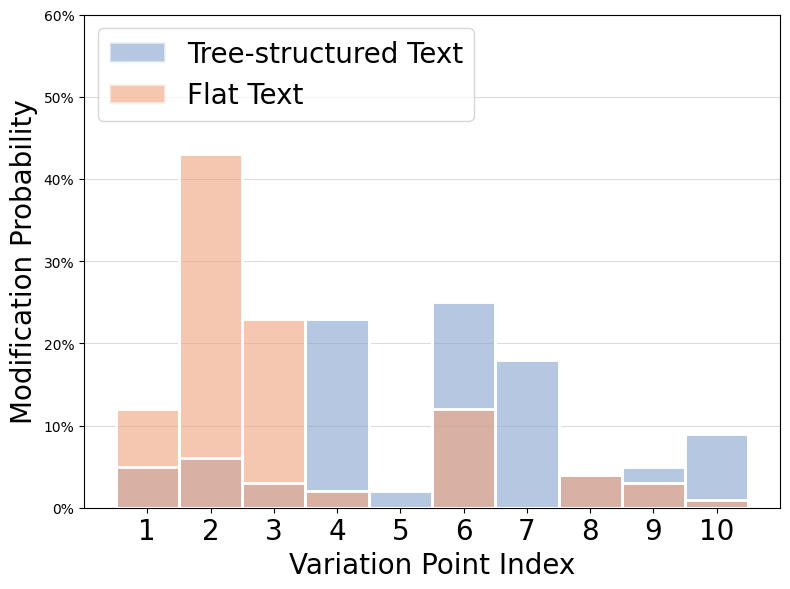}
        \label{fig:t2f0}
    \end{subfigure}
    \hfill
    \begin{subfigure}[b]{0.245\textwidth}
        \centering
        \includegraphics[width=\textwidth]{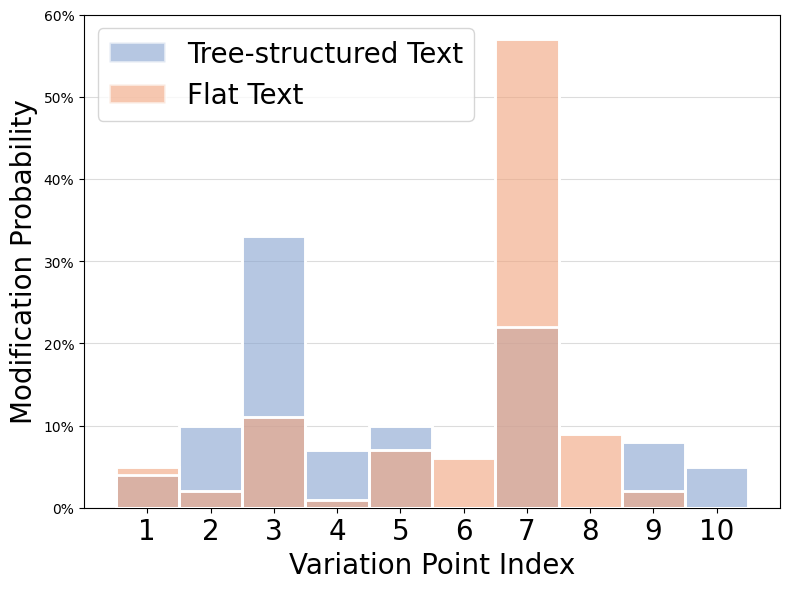}
        \label{fig:t2f1}
    \end{subfigure}
    \hfill
    \begin{subfigure}[b]{0.245\textwidth}
        \centering
        \includegraphics[width=\textwidth]{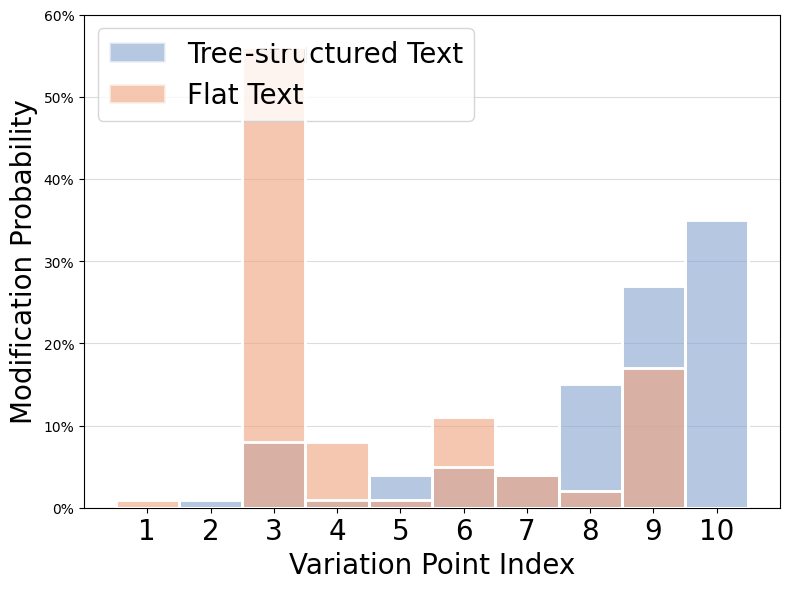}
        \label{fig:t2f2}
    \end{subfigure}
    \hfill
    \begin{subfigure}[b]{0.245\textwidth}
        \centering
        \includegraphics[width=\textwidth]{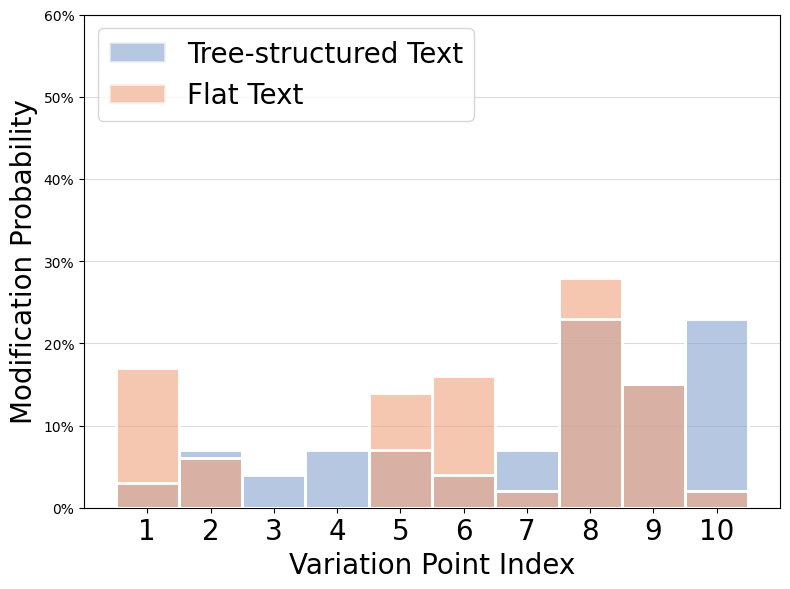}
        \label{fig:t2f3}
    \end{subfigure}
    \hfill
    \begin{subfigure}[b]{0.245\textwidth}
        \centering
        \includegraphics[width=\textwidth]{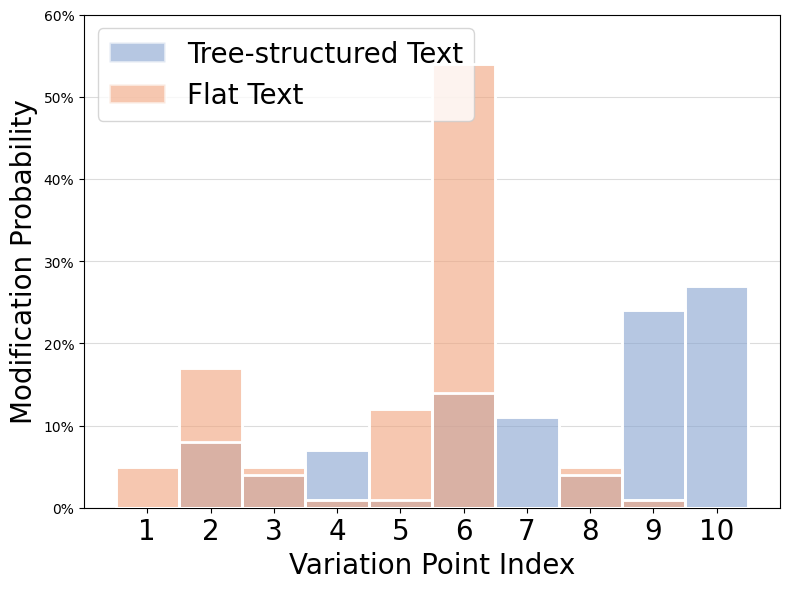}
        \label{fig:t2f4}
    \end{subfigure}
    \hfill
    \begin{subfigure}[b]{0.245\textwidth}
        \centering
        \includegraphics[width=\textwidth]{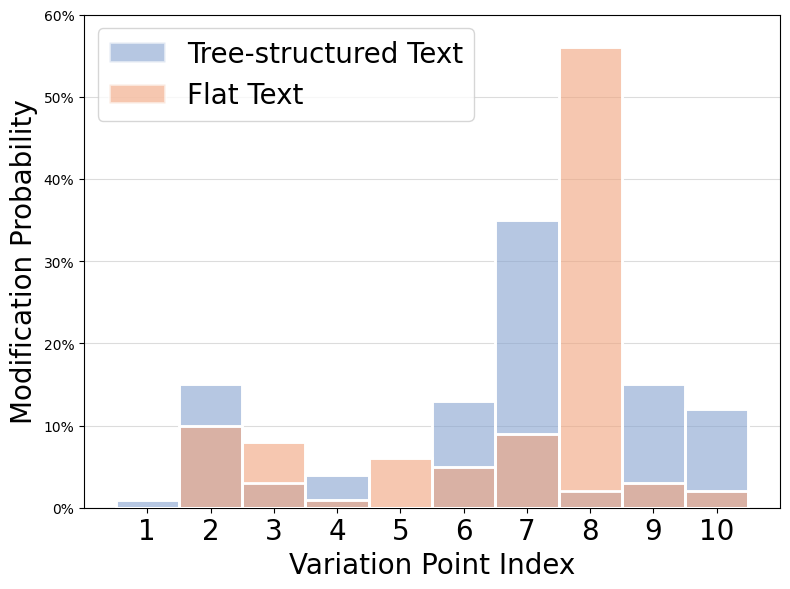}
        \label{fig:f2t0}
    \end{subfigure}
    \hfill
    \begin{subfigure}[b]{0.245\textwidth}
        \centering
        \includegraphics[width=\textwidth]{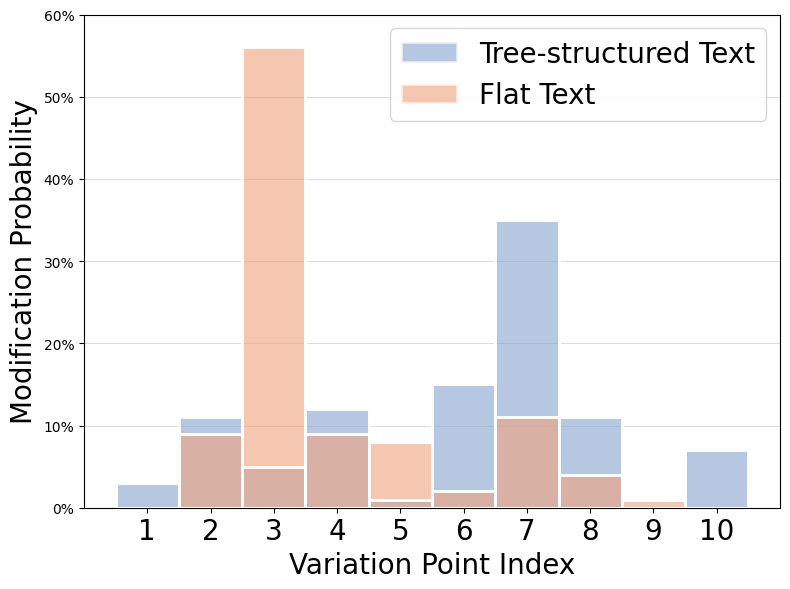}
        \label{fig:f2t1}
    \end{subfigure}
    \hfill
    \begin{subfigure}[b]{0.245\textwidth}
        \centering
        \includegraphics[width=\textwidth]{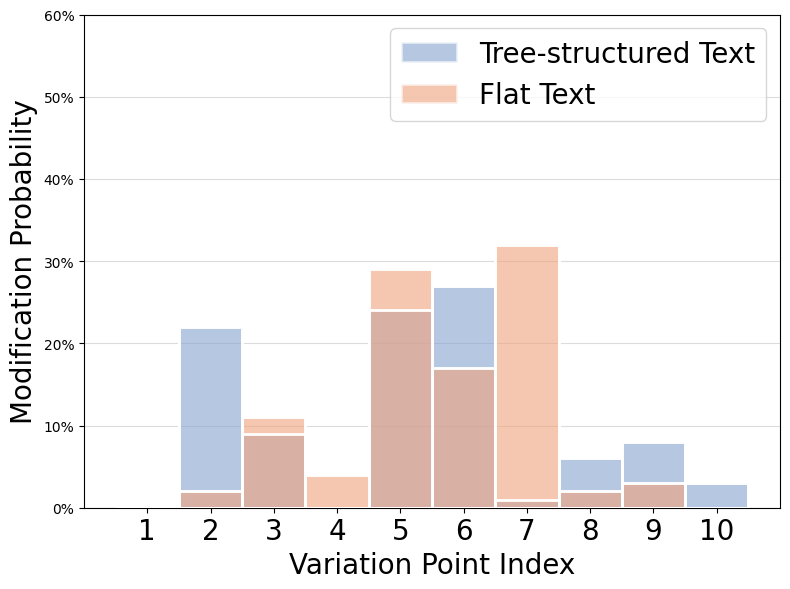}
        \label{fig:f2t2}
    \end{subfigure}
    \hfill
    \begin{subfigure}[b]{0.245\textwidth}
        \centering
        \includegraphics[width=\textwidth]{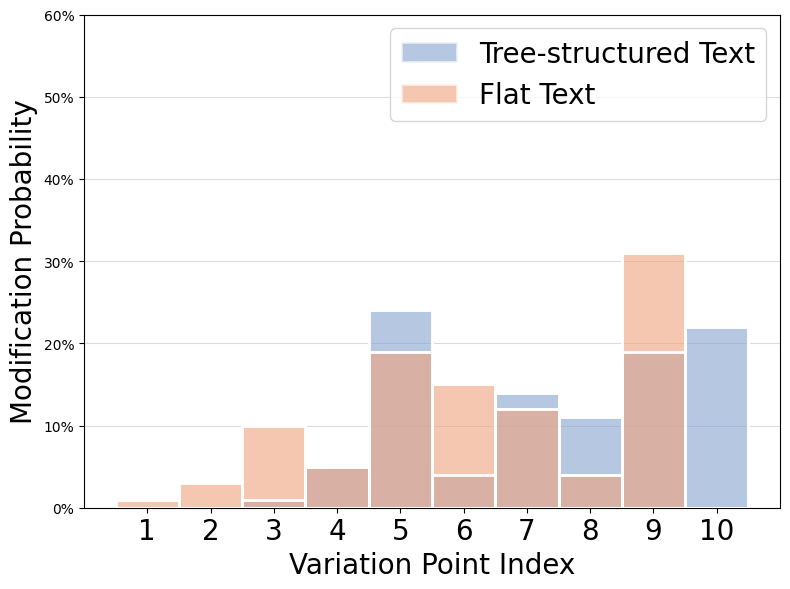}
        \label{fig:f2t3}
    \end{subfigure}
    \begin{subfigure}[b]{0.245\textwidth}
        \centering
        \includegraphics[width=\textwidth]{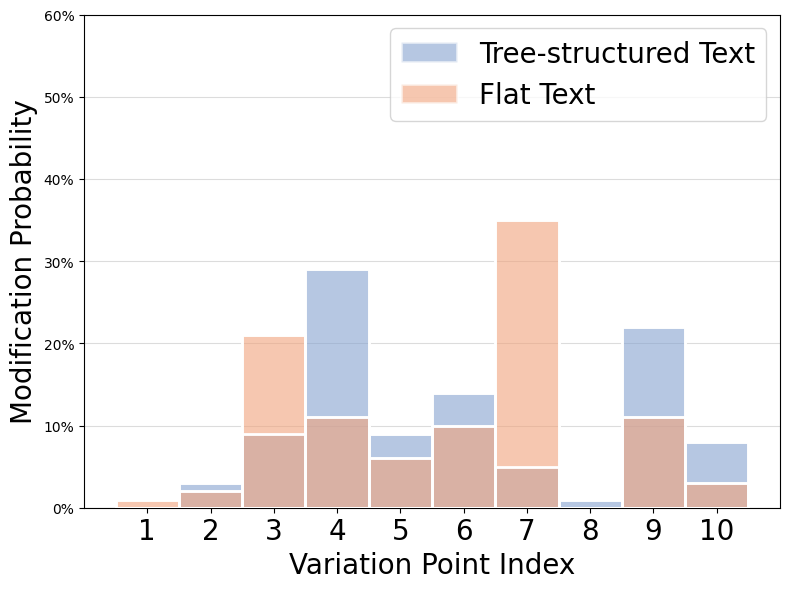}
        \label{fig:f2t4}
    \end{subfigure}
    
    \caption{Visualization of the remaining variation point distribution. For the first five images, The tree-structured text is generated and then transformed to flat text. For the last five images, The flat text is generated and then transformed to tree-structured text.}
    \label{fig:t2f1x5}
\end{figure*}

Figure \ref{fig:t2f1x5} shows the variation point distribution of 10 samples.

\subsection{Wall-Clock Time and Token Cost}
\label{app:clock}

We compare the Wall-Clock time and token cost under the experimental settings in our work. The result is provided in Figure \ref{tab:cost_comparison}. While LLM-based approaches costs less time than RL-based approaches, GP exhibits the lowest time cost. We further execute GP with a time budget of 20 min for 5 times, achieving an average IC of $0.0241\pm 0.0021$. Under the same time budget, TreEvo still outperforms GP.

\begin{table}[htbp]
\footnotesize
\centering
\caption{Wall-clock time and token cost comparison across algorithms.}
\label{tab:cost_comparison}

\begin{tabular}{lcc}
\hline
Algorithms & Wall-Clock Time & Token Cost \\
\toprule

GP         & 10 min   & -- \\
AlphaGen   & 4 hour   & -- \\
AlphaForge & 6 hour   & -- \\
QFR        & 4.5 hour & -- \\
TreEvo     & 20 min   & 654K \\
ReEvo      & 18 min   & 414K \\
EoH        & 40 min   & 632K \\
\bottomrule

\end{tabular}
\end{table}

\subsection{Factor Correlations}
\label{app:corr}

\begin{figure}[!htbp]
  \centering
  \begin{subfigure}[b]{0.45\linewidth}
    \centering
    \includegraphics[width=\linewidth]{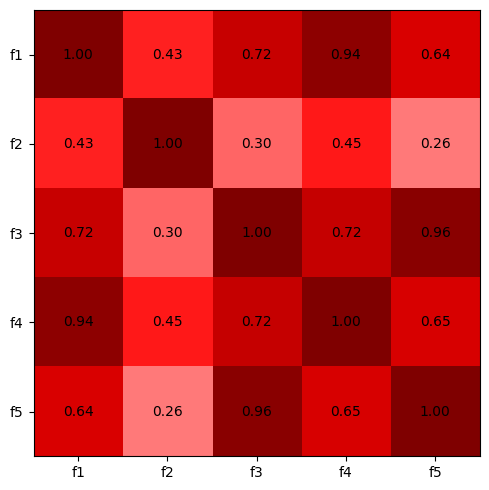}
    \caption{ReEvo}
    \label{fig:reevo_corr}
  \end{subfigure}
  \hspace{5pt}
  \begin{subfigure}[b]{0.45\linewidth}
    \centering
    \includegraphics[width=\linewidth]{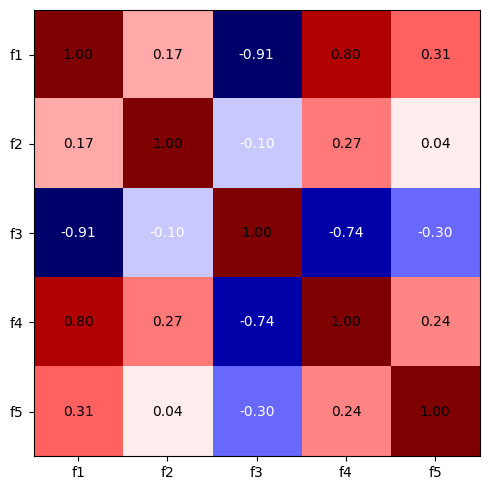}
    \caption{EoH}
    \label{fig:eoh_corr}
  \end{subfigure}

  \vspace{1em}

  \begin{subfigure}[b]{0.6\linewidth}
    \centering
    \includegraphics[width=\linewidth]{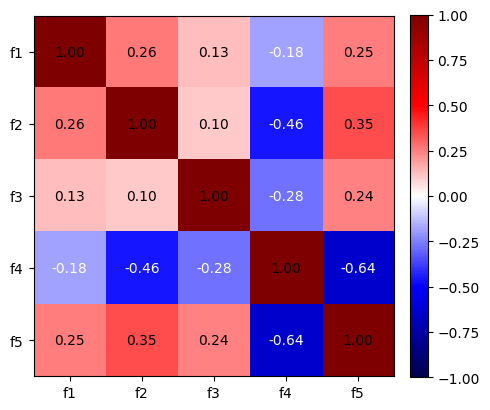}
    \caption{TreEvo}
    \label{fig:treevo_corr}
  \end{subfigure}

  \caption{IC correlation analysis of the best factors from 5 independent runs for LLM-driven EAs on CSI300.}
  \label{fig:corr}
\end{figure}

Figure \ref{fig:corr} shows the correlation among the best factors discovered by each method.

\subsection{Comparison with LLM Alpha Mining Frameworks}
\label{app:llm_frame}

Table \ref{tab:model_comparion} exhibits the performance of TreEvo compared to recent LLM alpha mining frameworks.

\begin{table}[htbp]
\footnotesize
\centering
\caption{Performance comparison across different markets with LLM alpha mining frameworks.}
\label{tab:model_comparion}

\begin{tabular}{llccc}
\hline
Dataset & Metric & RD-Agent & AlphaAgent & TreEvo \\
\toprule

\multirow{2}{*}{csi300}
& IC
& 0.0267
& 0.0278
& \textbf{0.0308} \\

& RankIC
& 0.0315
& 0.0329
& \textbf{0.0349} \\
\midrule

\multirow{2}{*}{csi500}
& IC
& 0.0283
& 0.0341
& \textbf{0.0362} \\

& RankIC
& 0.0344
& 0.0387
& \textbf{0.0393} \\
\midrule

\multirow{2}{*}{SPX}
& IC
& 0.0249
& \textbf{0.0323}
& 0.0317 \\

& RankIC
& 0.0286
& \textbf{0.0364}
& 0.0355 \\
\midrule

\multirow{2}{*}{NDX}
& IC
& 0.0214
& 0.0258
& \textbf{0.0285} \\

& RankIC
& 0.0263
& 0.0302
& \textbf{0.0316} \\
\bottomrule

\end{tabular}
\end{table}

\subsection{Comparison on Different LLMs}
\label{app:llm_perform}

Table \ref{tab:llm_comparion} exhibits the performance of TreEvo on different LLMs.

\begin{table*}[htbp]
\footnotesize
\centering
\caption{Performance comparison of TreEvo on different LLMs.}
\label{tab:llm_comparion}

\begin{tabular}{lcccccccc}
\midrule
LLM
& \multicolumn{2}{c}{csi300}
& \multicolumn{2}{c}{csi500}
& \multicolumn{2}{c}{SPX}
& \multicolumn{2}{c}{NDX} \\
\cmidrule{2-9}

& IC & RankIC
& IC & RankIC
& IC & RankIC
& IC & RankIC \\
\toprule

Qwen3-Max
& 0.0308 & 0.0349
& 0.0362 & 0.0393
& 0.0317 & 0.0355
& 0.0285 & 0.0316 \\

DeepSeek V3
& 0.0287 & 0.0337
& 0.0346 & 0.0375
& 0.0302 & 0.0347
& 0.0289 & 0.0331 \\

Gemini3 pro
& 0.0315 & \textbf{0.0368}
& \textbf{0.0378} & \textbf{0.0416}
& \textbf{0.0334} & \textbf{0.0371}
& \textbf{0.0327} & \textbf{0.0363} \\

GPT5.1
& \textbf{0.0318} & 0.0342
& 0.0358 & 0.0405
& 0.0323 & 0.0355
& 0.0309 & 0.0340 \\

\bottomrule
\end{tabular}
\end{table*}

\subsection{Walk-forward Validation}
\label{app:forward}

\begin{table*}[htbp]
\footnotesize
\centering
\caption{Quarterly performance comparison across methods.}
\label{tab:forward_validation}

\begin{tabular}{lccccccccc}
\hline
Quarter
& \multicolumn{3}{c}{EoH}
& \multicolumn{3}{c}{ReEvo}
& \multicolumn{3}{c}{TreEvo} \\
\cmidrule{2-10}

& IC & RankIC & ER
& IC & RankIC & ER
& IC & RankIC & ER \\
\toprule

Q1
& 0.0253 & 0.0319 & 0.0306
& 0.0228 & 0.0256 & 0.0144
& \textbf{0.0354} & \textbf{0.0392} & \textbf{0.0427} \\

Q2
& 0.0232 & 0.0259 & 0.0137
& 0.0227 & 0.0269 & \textbf{0.0189}
& \textbf{0.0278} & \textbf{0.0325} & 0.0158 \\

Q3
& 0.0273 & 0.0324 & 0.0301
& 0.0218 & 0.0261 & 0.0105
& \textbf{0.0343} & \textbf{0.0384} & \textbf{0.0431} \\

Q4
& 0.0249 & 0.0284 & 0.0223
& 0.0205 & 0.0245 & 0.0067
& \textbf{0.0326} & \textbf{0.0359} & \textbf{0.0356} \\

\bottomrule
\end{tabular}
\end{table*}

We apply walk-forward validation on CSI 300 by splitting the 2023 test period into quarterly segments (Q1–Q4), training each time on the preceding one year of data. The results, reported in terms of IC, RankIC, and excess return (ER), are shown in Table \ref{tab:forward_validation}.

\section{Framework and Prompts}
\label{app:prompt}

\subsection{Framework}

\noindent\textbf{Step 1 Initialization:} Initialize a population $P$ of $N$ thoughts by prompting LLM with the initialization prompt. Then ask LLM to generate $N$ codes based the $N$ thoughts and carry out the evaluation of each code respectively.

\noindent\textbf{Step 2 Evolutionary Process:} Repeat until the predefined stop criteria are met. 

    \noindent\textit{\textbf{Step 2.1 Offspring Generation:}} Generate a new offspring population $O$ of $N$ thoughts based on the current population $P$. For each iteration, the LLM is prompted to apply one of the following three operator types in rotation to generate all $N$ offspring:
    
    \begin{itemize}
        \item \textbf{Crossover:} Select two parent thoughts from $P$ and apply crossover operator to generate a new thought.
        \item \textbf{Mutation:} Select one parent thought from $P$ and apply Mutation-R, Mutation-I, or Mutation-F operator based on specified probabilities \(p_R\), \(p_I\), and \(p_F\) to generate a new thought.
        \item \textbf{Pruning:} Select one parent thought from $P$ and apply pruning operator to generate a new thought.
    \end{itemize}
    
    \noindent\textit{\textbf{Step 2.2 Evaluation:}} For each of the $N$ new offspring thoughts in $O$, use the LLM to generate the corresponding executable code. Evaluate each piece of code in the market dataset.

    \noindent\textit{\textbf{Step 2.3 Selection:}} Combine the $N$ parents from $P$ and the $N$ offspring from $O$ into a candidate pool of $2N$ individuals. Select the $N$ best-performing individuals from this pool to form the population $P$ for the next iteration.

\noindent\textbf{Step 3 Output:} Output the code of the best performing thought through the whole process.

\subsection{Prompts}

\begin{tcolorbox}[colback=white,colframe=black,title=System Prompt, breakable]
You are a “Quantitative Factor Tree Semantic Generator”.

Your task:
Randomly generate a “quantitative factor tree text” that has financial meaning, statistical logic, and computational validity.

Goal: \\
The generated result must resemble: \\
- a real quantitative factor \\
- an executable computation process \\
- a hierarchical semantic tree structure \\
- an AST (Abstract Syntax Tree) that progressively decomposes financial semantics into underlying data

-----------------------------------

[Output Format Requirements]

-----------------------------------

1. You must use ASCII tree format

2. The entire output must be a single tree

3. Output only the tree \\
Do not explain \\
Do not analyze \\
Do not add extra text \\

4. The root node must be: \\
an abstract financial objective, for example:

- Momentum Strength \\
- Volatility Compression \\
- Liquidity Imbalance \\
- Trend Persistence \\
- Earnings Quality \\
- Price Deviation \\
- Relative Strength \\
- Return Stability \\
- Volume Anomaly \\
- Market Sentiment Shift \\

5. The tree must satisfy:

abstract financial semantics \\
- intermediate statistical transformations \\
- underlying raw data \\

-----------------------------------

[Node Type Constraints]

-----------------------------------

Nodes may only belong to the following categories:

1. Financial semantic nodes \\
For example:

- Momentum \\
- Trend \\
- Volatility \\
- Liquidity \\
- Earnings Quality \\
- Valuation Deviation \\
- Sentiment Strength \\
- Relative Price Strength \\
- Capital Flow Pressure \\
- Microstructure Imbalance \\

2. Statistical / mathematical operation nodes \\
For example:

- addition \\
- subtraction \\
- ratio \\
- difference \\
- rolling$\_$mean \\
- rolling$\_$std \\
- rolling$\_$rank \\
- zscore \\
- normalize \\
- decay \\
- correlation \\
- covariance \\
- regression$\_$beta \\
- EMA \\
- rank \\
- abs \\
- log \\
- delta \\

3. Data leaf nodes \\
These must ultimately map to real financial data, including only:

- close \\
- open \\
- high \\
- low \\
- amount \\
- volume \\

-----------------------------------

[Structural Constraints]

-----------------------------------

1. Tree depth constraints:

Minimum depth: 3 \\
Maximum depth: 6

2. Number of child nodes per node:

- unary operations: 1 child \\
- binary operations: 2 children \\
- a small number of nodes may have 3 children

3. Degenerate chain structures are forbidden

4. Overly wide structures are forbidden

5. Every layer must have a clear semantic hierarchy:

higher levels are more abstract \\
lower levels are more concrete

-----------------------------------

[Financial Semantic Consistency]

-----------------------------------

1. Volatility-related nodes: \\
may only be composed from:

- rolling$\_$std \\
- variance \\
- abs(return) \\
- squared$\_$return \\

2. Momentum-related nodes: \\
may only be composed from:

- return \\
- ratio \\
- delta \\
- EMA trend \\
- cumulative return

3. Liquidity-related nodes: \\
may only use the following data:

- volume \\
- turnover \\
- amount \\
- spread

4. Node names must be logically consistent with their subtrees

-----------------------------------

[Time Constraints]

-----------------------------------

1. Future information is forbidden

Forbidden:

- return$\_${t+1} \\
- close$\_${t+1}

2. Allowed:

- t \\
- t-k \\
- rolling window

-----------------------------------

[Generation Requirements]

-----------------------------------

1. Each generation must be random

Randomization includes:

- financial theme \\
- operation combinations \\
- time windows \\
- data sources \\
- normalization methods \\
- tree structure shape

2. Do not generate repeated templates

3. Do not generate meaningless stacks of financial buzzwords

Incorrect:

market volatility trend liquidity strength factor

4. The tree must be mappable to a real computational expression
\end{tcolorbox}

\begin{tcolorbox}[colback=white,colframe=black,title=Initialization Prompt, breakable]
$\{seed\_tree$\} \\

Refering to the format of a trivial design above, be very creative to generate a totally different factor from the given factor and give the thought tree of it. Your response only outputs txt code of the thought tree. Format your code as a txt code block: ```txt ... ```.
\end{tcolorbox}

\begin{tcolorbox}[colback=white,colframe=black,title=Crossover Prompt, breakable]
[Crossover]
Crossover generates a new factor tree by refering the high-level ideas from parent trees. Choose a sub-idea from one parent tree and combine it with the other parent tree. You are allowed to revise the high level ideas of changing sub-idea to preserve logical consistency and remain interpretable. \\

[Parent tree 1] \\
$\{parent\_tree\_1\}$ \\

[Parent tree 2] \\
$\{parent\_tree\_2\}$ \\

[New tree] \\
The root should represent the shared main idea. Child nodes should be meaningful refinements of their parent. The remaining branches should be limited unchanged. Your response only outputs txt code of the thought tree. Format your code as a txt code block: ```txt ... ```.
\end{tcolorbox}

\begin{tcolorbox}[colback=white,colframe=black,title=Mutation-R Prompt, breakable]
[Mutation]
Mutation generate the tree of the new factor which is totally different from all parent trees to ensure diversity. The new tree should be logical and interpretable. \\

[Parent trees] \\
$\{parent\_trees\}$ \\

[New tree] \\
Generate a single coherent tree structure. Your response only outputs txt code of the thought tree. Format your code as a txt code block: ```txt ... ```.
\end{tcolorbox}

\begin{tcolorbox}[colback=white,colframe=black,title=Mutation-I Prompt, breakable]
[Sub-idea Mutation]
Sub-idea Mutation generate a new tree by changing one sub-idea of the parent tree. First, choose one internal node of the tree. Then introduce a new sub-idea to replace the old sub-idea while keeping logical consistency of the tree. The new tree should be logical and interpretable. \\

[Parent tree] \\
$\{parent\_tree\}$ \\

[New tree] \\
Generate a single coherent tree structure. Your response only outputs txt code of the thought tree. Format your code as a txt code block: ```txt ... ```.
\end{tcolorbox}

\begin{tcolorbox}[colback=white,colframe=black,title=Mutation-F Prompt, breakable]
[Parameter Mutation]
Parameter Mutation generate a new tree by changing the parameters in the parent tree while maintaining the idea of the tree. The Parameters exists all on leaf nodes in the tree. The new tree should be logical and interpretable. \\

[Parent tree] \\
$\{parent\_tree\}$ \\

[New tree] \\
Generate a single coherent tree structure. Your response only outputs txt code of the thought tree. Format your code as a txt code block: ```txt ... ```.
\end{tcolorbox}

\begin{tcolorbox}[colback=white,colframe=black,title=Pruning Prompt, breakable]
[Pruning]
Pruning simplifies a thought tree by keeping core idea and removing redundant or low-importance components. The main idea and overall structure should be preserved. Do not introduce any new concepts or rewrite existing ones. \\

[Parent tree] \\
$\{parent\_tree\}$ \\

[New tree] \\
Generate a single coherent tree structure. Each remaining node should directly support the core reasoning idea. Your response only outputs txt code of the thought tree. Format your code as a txt code block: ```txt ... ```.
\end{tcolorbox}

\end{document}